\documentclass[twocolumn,pre, superscriptaddress, showpacs]{revtex4}
\usepackage{hyperref,xcolor}
\usepackage{amsmath,amssymb}
\usepackage{graphicx}
\usepackage{xcolor}

\newcommand{\imag}{\ensuremath{\mathrm{i}}}
\renewcommand\Im{\ensuremath{\mathrm{Im}\,}}
\renewcommand\Re{\ensuremath{\mathrm{Re}\,}}

\newcommand\Dfrtl[1]{\ensuremath{\,\mathrm{d}#1\,}}
\renewcommand\epsilon{\varepsilon}

\newcommand\euler[1]{\ensuremath{\mathrm{e}^{#1}}}

\newcommand\omegaii{\gamma}

\begin{document}
\title{
Extracting spectral properties from Keldysh Green functions
}
\date{\today}

\author{Andreas \surname{Dirks}}
\affiliation{Department of Physics, University of G\"ottingen, D-37077 G\"ottingen, Germany}
\affiliation{Department of Physics, Georgetown University, Washington, DC, 20057 USA}
\author{Martin \surname{Eckstein}}
\affiliation{Max Planck Research Department for Structural Dynamics, University of Hamburg-CFEL, Hamburg, Germany}
\author{Thomas \surname{Pruschke}}
\affiliation{Department of Physics, University of G\"ottingen, D-37077
G\"ottingen, Germany}
\author{Philipp \surname{Werner}}
\affiliation{Department of Physics, University of Fribourg, 1700 Fribourg, Switzerland}

\begin{abstract}
We investigate the possibility to assist the numerically 
ill-posed calculation of spectral properties of interacting quantum systems
in thermal equilibrium 
by extending the imaginary-time 
simulation to a finite Schwinger-Keldysh contour.
The effect of this extension is tested within the standard Maximum Entropy 
approach to analytic continuation.
We find that the inclusion of real-time data improves the resolution of structures at high energy, while the imaginary-time data
are needed to correctly reproduce low-frequency features such as quasi-particle peaks. 
As a nonequilibrium application, we consider the calculation of
time-dependent spectral functions from retarded Green function data on a
finite time-interval, and compare the Maximum Entropy approach to direct
Fourier transformation and a method based on Pad\'e approximants. 
\end{abstract}


\maketitle

\section {Introduction}
A common strategy to circumvent the oscillatory convergence of integrals 
in interacting quantum field theories is the Wick rotation to 
imaginary time. Apart from possible fermionic sign problems, quantum 
Monte Carlo (QMC) simulations are well-conditioned on the imaginary 
time axis, and  simulations at non-zero temperature using 
the Matsubara technique are straightforward. 
However, extracting dynamic properties from imaginary-time  
data of the single-particle Green's function
$G(-\imag\tau)=-\langle \text{T}_\tau c(\tau) c^\dagger(0)\rangle$    
is an 
ill-posed problem: the singular integral equation
\begin{equation}
\imag G(-\imag\tau) = \int \Dfrtl{\omega} 
    A(\omega) \frac{\euler{-\tau \omega}} {1 + \euler {-\beta \omega}}
\label{eq:kernelMatsubara}
\end{equation}
has to be solved for the spectral function $A(\omega)$.

If the real-time data for the retarded Green's function
\begin{equation} 
G^\text{ret}(t,t')=-i\Theta(t-t')\langle\{d(t),d^\dagger(t')\}\rangle\label{retarded}
\end{equation}
were known on the entire 
real-time axis, the spectral function could be obtained from a simple inverse 
Fourier transform (in equilibrium, $G^\text{ret}(t,t')$ depends only on the time difference $t-t'$).
The Fourier transform is not ill-posed, due to
unitarity. Unfortunately, calculating these real-time data for an interacting system is difficult. Monte Carlo 
techniques cannot reach long times, due to a dynamical sign problem
which grows exponentially as a function of the maximum real time $t_\text{max}$.
Nevertheless, as shown in Refs.~\cite{Eckstein09, Eckstein10, Werner10} the real-time Green function 
up to some finite time $t_\text{max}$ can be computed accurately using continuous-time Monte Carlo 
algorithms \cite{Rubtsov05, Werner06, Muehlbacher08, Werner09}, or, in certain parameter regimes, 
using perturbative weak- \cite{Tsuji2012afm} or strong-coupling methods \cite{Gull2010, Eckstein10nca}. This raises the question if the 
information contained in the real-time correlators for $t, t'<t_\text{max}$ can be exploited to obtain reliable spectra
using a suitably adapted analytical continuation procedure. 

A real need for the ``analytical continuation" of 
real-time Green functions   
arises in the study of nonequilibrium properties of correlated systems, e.~g. lattice systems perturbed by a quench or some field pulse \cite{Eckstein2009, Eckstein2011pump, Tsuji2012}. In order to characterize the relaxation of these systems, it is often useful to define a time-dependent spectral function
\begin{equation}
A(\omega,t)=-\frac{1}{\pi}\text{Im}\int_t^\infty dt' e^{i\omega(t'-t)}G^\text{ret}(t',t),
\label{def_a_t}
\end{equation}
where $G^\text{ret}(t',t)$ now depends on both times individually due to the loss of time-translation invariance.
$A(\omega,t)$ is not assured to be
positive, but it typically becomes positive a short time after the
perturbation. (In particular,  $A(\omega,t)$ is positive for any
quasi-stationary state. When $A(\omega,t)$ is constant
over a time window of width $\Delta t$, then $A(\omega,t)$ must be positive 
after averaging over a frequency window of $\Delta \omega \propto 1/\Delta t$.)
Under certain assumptions,   $A(\omega,t)$ is related 
to the photoemission and inverse photoemission signal \cite{Eckstein2008, Freericks2009}. 
Furthermore, in
an equilibrium system, Eq.~(\ref{def_a_t}) reduces to the familiar spectral
function defined in Eq.~(\ref{eq:kernelMatsubara}). The challenges in the
evaluation of $A(\omega,t)$ are the same as those for equilibrium spectra
mentioned above: In practice, simulation results will be limited in time, so
direct Fourier transformation will lead to oscillations or a smearing-out of
spectral features. Therefore, a second question which we want to address is whether 
Maximum Entropy or Pad\'e approaches can be used to improve the quality of these time-dependent spectra.

The paper is structured as follows: in Sec.~\ref{MEM} we present and test the 
Maximum Entropy approach for Keldysh Green's 
functions. In particular, 
Sec.~\ref{singvals} investigates 
the singular values of the kernel matrix and their dependence on the choice of
data 
points,  in 
Sec.~\ref{artificial} we test the analytical continuation
procedure for exactly known Green's 
functions, and 
in Sec.~\ref{inversion} we compare the Maximum Entropy result to spectra obtained by Fourier transformation. 
 A generalization of the Pad\'e analytical continuation procedure to nonequilibrium Green functions is introduced in Sec.~\ref{Pade}. In Sec.~\ref{application} we apply the 
Maximum Entropy method to real-time simulation data of equilibrium and
nonequilibrium systems, and compare the method to direct Fourier
transformation and the Pad\'e procedure. Sec.~\ref{conclusion} 
gives 
a summary and conclusion.

\begin{figure}
\begin{centering}
\includegraphics[width=\linewidth]{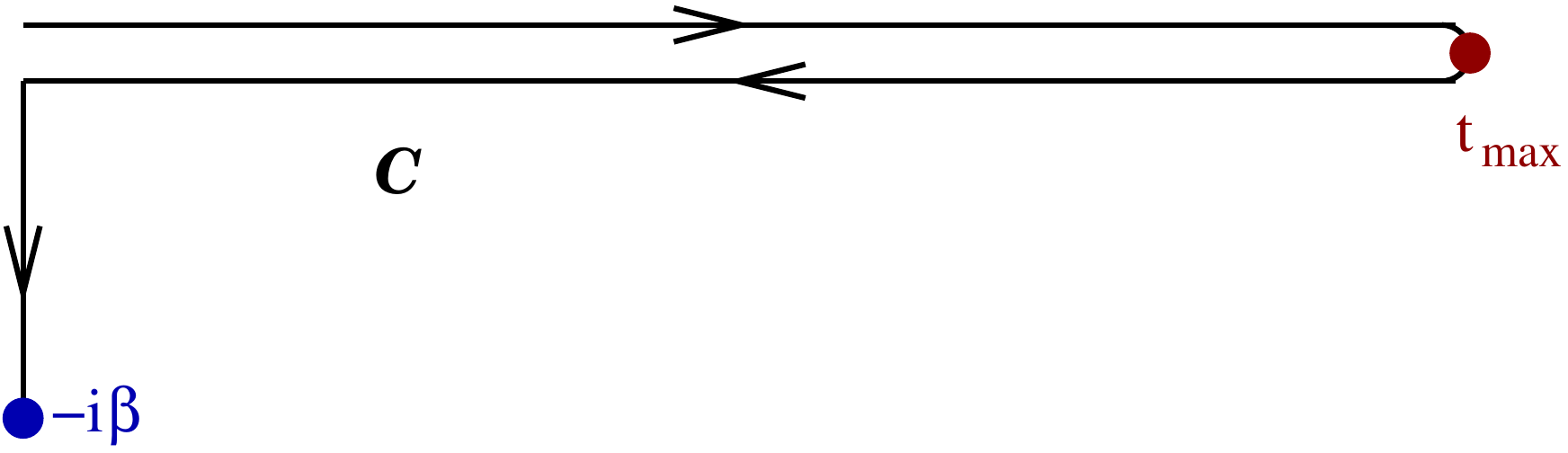}
\end{centering}
\caption{(color online) The Schwinger-Keldysh contour $\mathcal{C}$. Arrows denote the direction of
time-ordering $<_\mathcal{C}$. 
}
\label{fig:contour}
\end{figure}

\section{Maximum Entropy approach}

\label{MEM}

\subsection{Formulation of the problem}

\label{kernel}

In practice, real-time simulations are carried out on the Keldysh contour 
$\mathcal{C}$ 
\cite{Schwinger61,keldyshintro}  
illustrated in Fig. \ref{fig:contour}. The Green's function 
is a function of two time indices $t$ and $t'$, which may lie on the upper real, the lower real, 
or the Matsubara branch of the contour.
Using the Keldysh-contour time-ordering $<_\mathcal{C}$, the 
contour-ordered 
equilibrium Green's function 
is related to the spectral function through 
\begin{align}
\imag G(z,z') &= 
\langle \text{T}_\mathcal{C}d(z)d^\dagger(z')\rangle \nonumber
\\
&= \int \Dfrtl{\omega}
A(\omega) \euler{-\imag (z-z') \omega}
F_{z,z'}(\omega),
\label{eq:kernelKeldysh}
\end{align}
with the contour-ordered Fermi factor 
\begin{equation}
F_{z,z'}(\omega):=
\begin{cases}
f(-\omega) & \text{if } z' <_\mathcal{C} z, \\
-f(\omega) & \text{else},
\end{cases}
\end{equation}
and the times $z$, $z'$ located on $\mathcal{C}$.
Obviously, Eq.~\eqref{eq:kernelKeldysh} is a generalization of 
Eq.~\eqref{eq:kernelMatsubara},
and extracting $A(\omega)$ from any data set $G(z_i,z_j)$
is still an ill-posed problem when the times $z_i$ are restricted to
the Matsubara branch, or to times smaller than some finite 
$t_\text{max}$.
 
The Maximum Entropy 
analytic continuation procedure 
involves the inference of a \emph{most
probable} spectral function $A(\omega)$ among those which are compatible with the
simulated data. 
The problem can be expressed in the matrix form 
\begin{equation}
\label{kernel-def}
D = K \cdot A,
\end{equation}
where $D$ represents simulation data, $K$ is a linear operator, given by the matrix
kernel, applied to the unknown spectral function $A$. The finite set $D$ can
be interpreted as a snapshot from a Gaussian random variable with covariance 
$C$, in the case of a Monte Carlo simulation \cite{jarrell_review}. 
\par
The essential idea for inferring a most probable $A$ from $D$ has been outlined 
in Ref.~\cite{jarrell_review}. An entropy functional $\alpha S[A]$ is added in
order to regularize the 
minimization of $\chi^2[D,A]$ with respect to $A$,
where 
\begin{equation}
\chi^2[D,A] := \frac{1}{2}(KA-D)^TC^{-1}(KA - D).
\end{equation}
The most
informative scalar $\alpha$ can be determined by a systematic application of 
Bayesian inference (\emph{Classic MaxEnt}), or can be averaged over the
corresponding Bayesian probability distribution (\emph{Bryan's MaxEnt}).

The standard numerical algorithm for the 
maximum entropy 
inference of spectral functions 
was developed by Bryan \cite{bryan}. The kernel is decomposed via a
singular-value decomposition
\begin{equation}
K = V \Sigma U^T,
\end{equation}
where $\Sigma = \mathrm{diag}\,(\sigma_1, \sigma_2,\,\dots)$,
$\sigma_1\geq\sigma_2\geq \cdots \geq 0$. 
Through the matrix products, each singular value $\sigma_i$ is associated with a direction
in $A$-space and a direction in $D$-space. The former is given by one of the ``basis functions" for the spectral function
and the latter is represented by an entry of $V$.

If $\sigma_i$ is 
large, it provides a channel which transports a comparably large amount of
information about $A$. If $\sigma_i$ is small, not much information can be
gained for the corresponding direction in $A$-space, and the Bayesian
approach will not modify a default spectrum with respect to that direction, due to a lack
of evidence. 
A small value of $\sigma$ can only be compensated by small 
error bars for the corresponding $D$-direction. 
Hence, the shape of the singular value distribution is an important indicator for the
structure of the inverse problem.

\subsection{Singular value distributions}
\label{singvals}

\begin{figure}
\includegraphics[width=\linewidth]{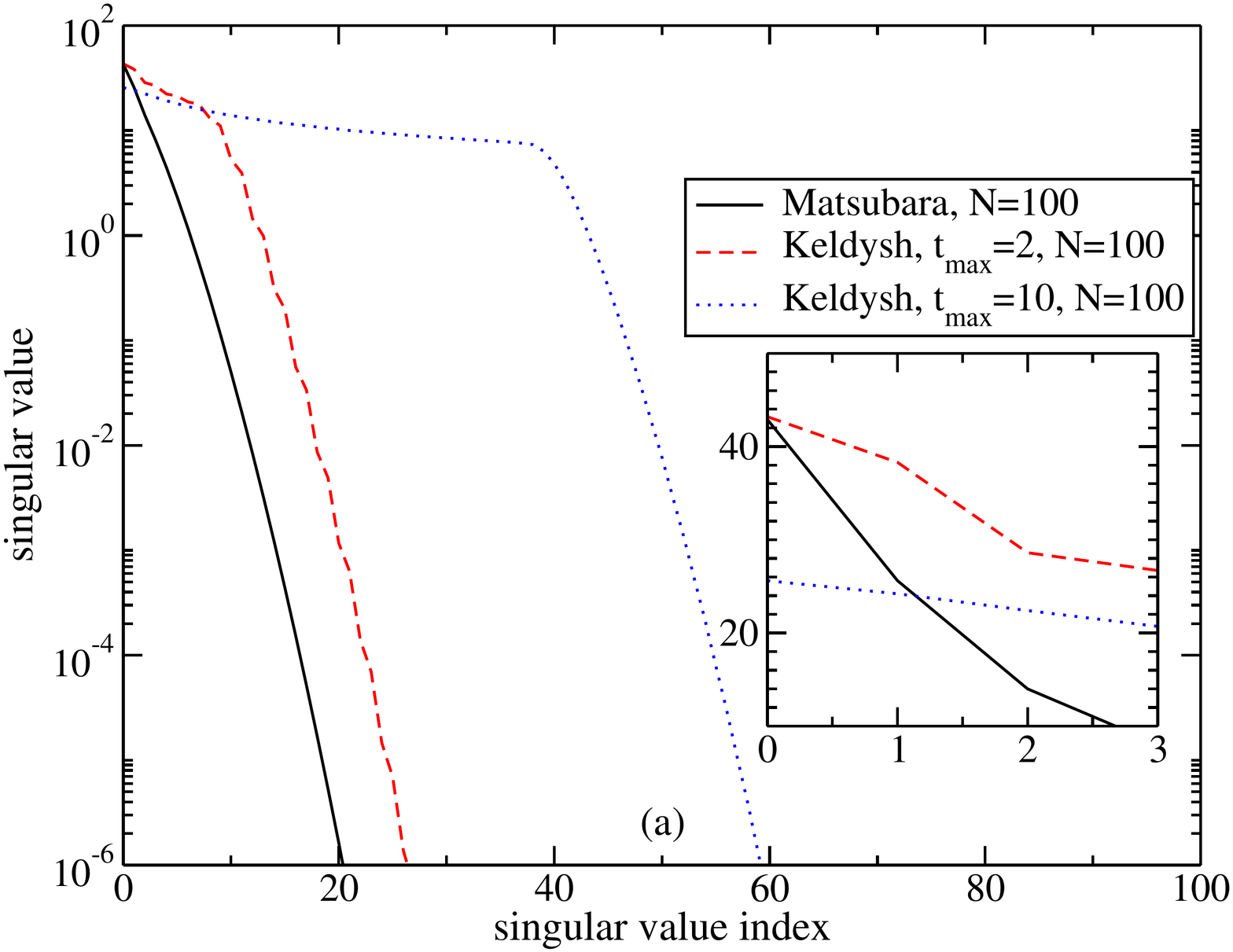}\\
\includegraphics[width=\linewidth]{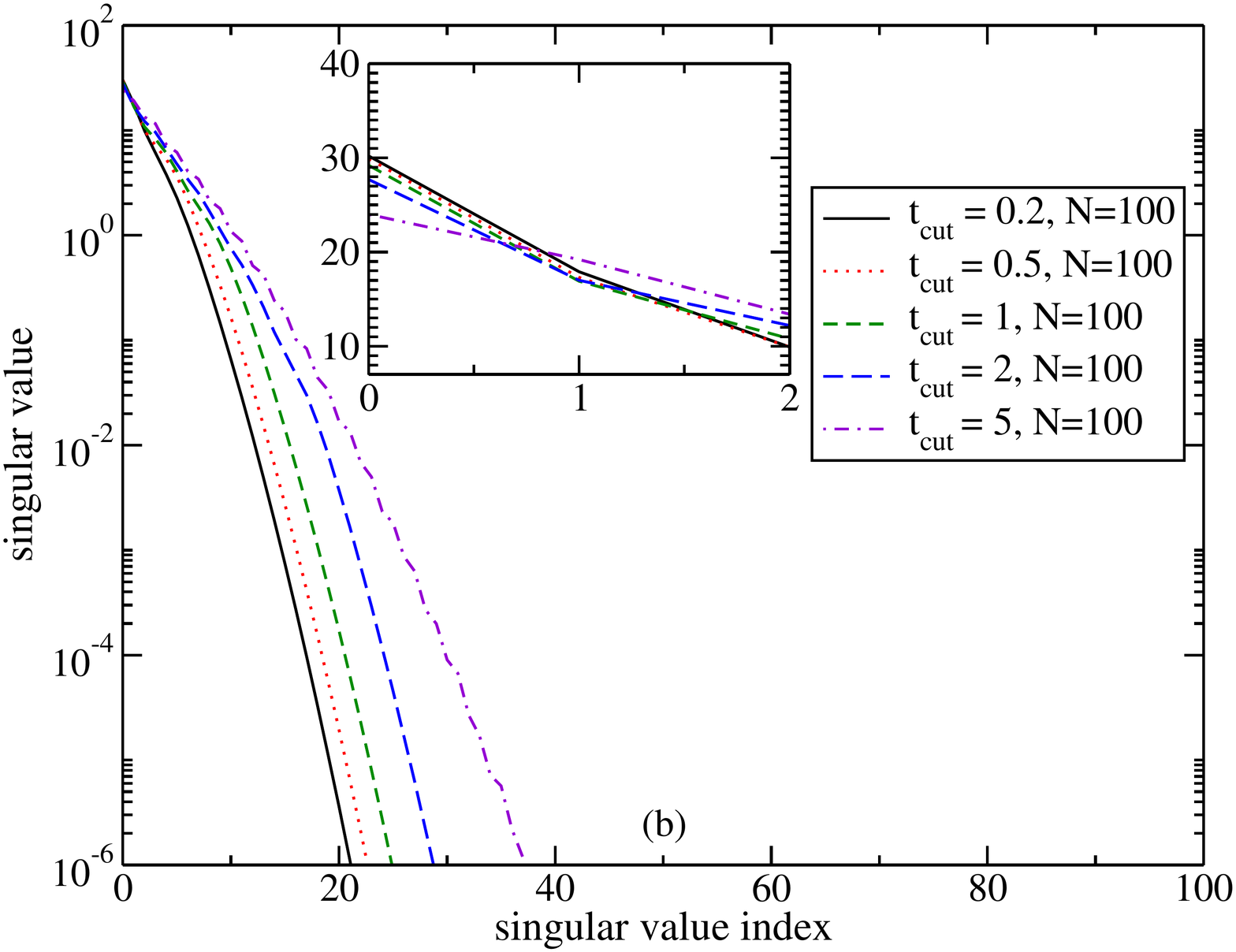}
\caption{(color online) Singular value distributions for $\beta=10$ and
indicated number of time-points $N$. Panel (a): Matsubara and real-time Green's functions. 
Panel (b): Keldysh-offdiagonal Green's functions.}
\label{fig:mixedsvs}
\label{fig:singvals}
\end{figure}

In this section we analyze the singular value structure of the integral kernel $K$ in 
Eq.~(\ref{kernel-def}), for various data sets $D$ (the set $A$ is always
given by some appropriate discretization of the spectral function). 
In panel (a) of Fig.~\ref{fig:singvals} the singular value distribution is plotted for the
usual inference from imaginary-time data (solid lines), for inverse temperature $\beta=10$. 
In this case, the data set $D=D_\text{imag}$  consists of the values $G(-i\tau_j,0)$ on $N$
equidistant points on the imaginary time branch,
\begin{equation}
D_\text{imag}^{N,\beta} = \{ G(-i \tau_j ,0) \,|\, \tau_j=\beta j /N,  j=0, \ldots,  N-1 \}.
\label{set-imag}
\end{equation}
The singular values are seen to decay exponentially.

Next we define a suitable data set for extracting $A(\omega)$ from pure real-time data. We note that 
the imaginary-time set of input data contains real-time information for both positive (forward direction) and
negative (backward direction) frequencies. In the case of real-time data, the situation is different: 
If only one time ordering of $d, d^{\dagger}$ is taken into account, the Maximum Entropy method 
only yields information about the positive or negative frequency range, i.~e. the occupied and 
unoccupied part of the spectrum.  It is thus important to consider  ``lesser" {\em and} ``greater"
(or retarded)
Green's 
functions. We choose $(z,z')$ values that correspond 
to both electrons propagating forward in time starting from $t'=0$, $G^>(t,0)=-i\langle d(t)d^\dagger(0)\rangle$,  
and backward in time starting from $t'=t_\text{max}$ along the upper real-time branch, $G^<(t,t_\text{max})=
i\langle d^\dagger(t_\text{max}) d(t)\rangle$, using an equidistant grid of annihilation times, 
\begin{align}
& D_\text{real}
^{N,t_\text{max}} =
\{ \text{Re} G^>(t_j,0),\text{Im} G^>(t_j,0),\text{Re} G^<(t_j,t_\text{max}),
\nonumber\\
&\hspace{7mm}\text{Im} G^<(t_j,t_\text{max})  \,|\,  t_j=jt_\text{max}4/N,  j=0, \ldots,  N/4-1 \}.
\label{set-real}
\end{align}
Due to translational invariance, the $d^\dagger$ operator need not be shifted in time.
Since we use both the real and imaginary parts of
$G^>(t,0)$ and $G^<(t,t_\text{max})$ 
each time step contributes two real variables to the inference process, and we always use $N$ 
for the the total number of real variables below.

In contrast to the case of imaginary-time input, the kernel for pure real-time propagators has a distinguished plateau in its singular 
value distribution, even for rather small $t_\text{max}=2$ (dashed line in Fig.~\ref{fig:singvals}).
As $t_\text{max}$ is increased ($t_\text{max}=10$, dotted line), the plateau broadens, and more equally dominant
$A$-directions appear. This is because in the limit $t_\text{max}\to\infty$ 
the (no longer ill-posed) unitary limit $\sigma_i \equiv 1$ is approached.

We next consider singular-value distributions for Keldysh-offdiagonal propagators.
For symmetry reasons we can restrict ourselves to mixed Green functions from the upper Keldysh contour to
the imaginary branch, $G^{\lceil}(\tau, t_\text{cut})$.
We consider fixed values of $t_\text{cut}$. $t_\text{cut}=0$ is equivalent to the usual
kernel for the Matsubara Green's function. As $t_\text{cut}$ is increased,
the singular value distribution is broadened (panel (b) of Fig.~\ref{fig:mixedsvs}), 
but not as dramatically as in panel (a) of Fig.~\ref{fig:mixedsvs}.
Keeping $\tau=\tau_\text{cut}$ fixed yields a structure more similar to the
real-time distributions in 
panel (a) ($\tau_\text{cut}=0$ is exactly the same). 
However, it appears that the plateaus in panel (a) cannot
be exceeded.

\begin{figure}
\includegraphics[width=\linewidth]{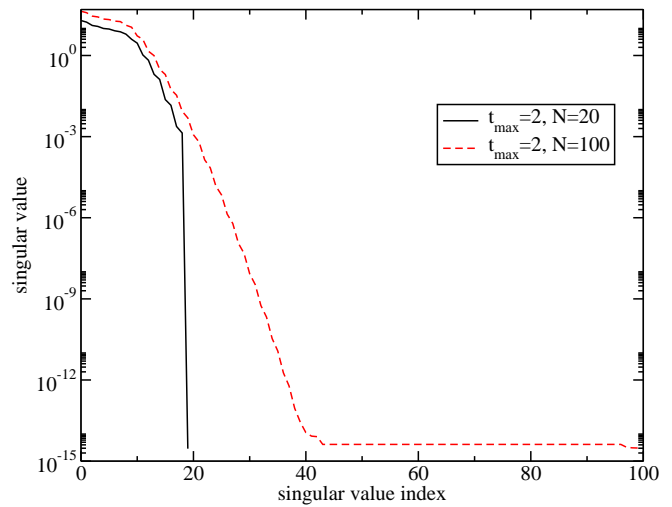}
\caption{(color online) Dependence of real-time singular value distributions on the
number of input data $N$ 
in $D_\text{real}^{N,t_\text{max}}$.
}
\label{fig:singvals_dependenceonN}
\end{figure}

Figure \ref{fig:singvals_dependenceonN} shows the dependence of the real-time
singular-value distribution on the number $N$ of time-points, for a Keldysh contour length $t_\text{max}=2$. We find that at
low singular value indices, apart from a finite offset, the same shape is
obtained. In particular, the same number of singular values is associated with 
the plateau. Furthermore, the initial descent from the plateau is also
identical. However, the singular value distribution for $N=100$ continues to decrease
smoothly, whereas the corresponding curve for $N=20$ abruptly jumps to the lowest singular value 
which is of the order of $10^{-15}$.
A similar behaviour is also found for other correlators within the Keldysh
contour.

We conclude that increasing $N$ does not extend the width of the plateau (once the latter has been established), but merely adds 
further singular values to the rapidly decreasing tail. Since this decrease is exponential, an increase in $N$ is only useful
if the added singular values are larger than the order of magnitude
$\epsilon$ of the numerical error of the data.
This dependence is indeed observed in the data analyis presented in the following subsections.

\subsection{Tests of equilibrium spectral functions}
\label{artificial}

We will first analyze the behaviour of the continuation procedure using artificial
data sets corresponding to 
given 
spectral functions with sharp peaks or band edges. For this purpose,
uncorrelated data sets taken from the exact contour Green function are 
studied, and 
we assume 
a uniform error distribution $\epsilon$, 
i.e., 
the covariance matrix is taken to be of the form 
$C = \text{diag}\,(\epsilon^2)$. The performance of the Maximum Entropy procedures on more
realistic data sets will be investigated in Sec.~\ref{application}.

\subsubsection{Rectangular spectrum}

As a first test case we consider the rectangular spectral function
\begin{equation}
A_\text{rect}(\omega) := \frac{1}{4}\cdot
\begin{cases}
1, & \text{if } |\omega|<2, \\
0, & \text{else}.
\end{cases}
\end{equation}
It can be expected that the sharp band edges will be difficult to infer from any finite data set.
It is well-known that even the inverse Fourier transform, i.e.~the unitary 
limit $t_\text{max}\to\infty$ converges only slowly, and that the convergence is not 
point-wise. Hence, we consider this an interesting test case for the Maximum
Entropy method.
The analytic continuation procedure is tested for inverse 
temperature $\beta = 10$ and for real-branch lengths $t_\text{max}=2$ and $t_\text{max}=10$.
The fake variance of all $\mathcal{C}$-contour correlator estimators is set 
to $\epsilon^2 = 6\cdot 10^{-14}$. 

\begin{figure}
\begin{center}
\includegraphics[width=\linewidth]{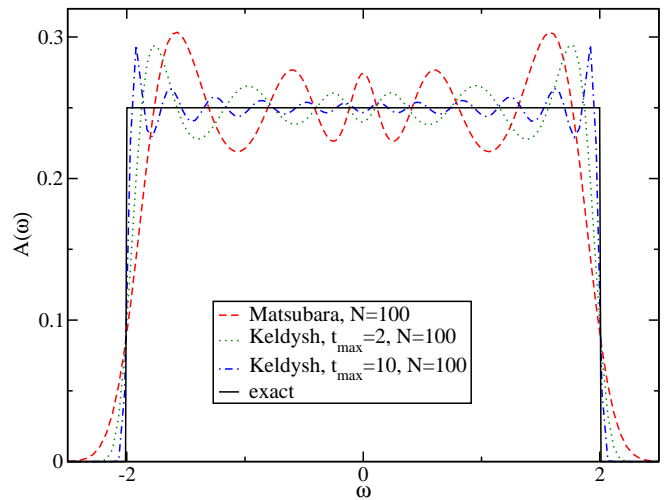}
\end{center}
\caption{(color online) Comparison of spectra at $\beta = 10$ for $N=100$ data points located on the
Matsubara contour, or on the upper Keldysh contour up to time $t_\text{max}=2$, $10$.
}
\label{fig:rect}
\end{figure}

Figure \ref{fig:rect} compares the exact spectral function to the $A(\omega)$ obtained from the analytical 
continuation for different 
data sets (\ref{set-imag}) and (\ref{set-real}).
The total number of real variables is kept constant at $N=100$, i.e., we use either 100 equidistant time steps 
on the imaginary axis, or 25 time steps in the real-time set $D_\text{real}^{N,t_\text{max}}$.
(We found that the restriction of real  input data to the real or imaginary parts of $G(t,t')$ 
does not contain sufficient information for an analytical continuation.)

In Fig.~\ref{fig:rect} it can be seen that using a broad Gaussian default
model, the width of which has no significant influence on the results, the 
maximum entropy 
solution shows Gibbs ringing artifacts. The frequency of these 
oscillations is a measure of the accuracy of the inferred spectrum.
Surprisingly, already the real-time data from a short Keldysh branch,  $t_\text{max}=2$, 
yield more accurate results than the $\beta=10$ Matsubara branch data. 
As $t_\text{max}$ is increased, the approximation of the spectrum becomes even better.

Further increasing the number of data points $N$ does not significantly change the above results. 
However, lowering the variance $\epsilon^2$ and then raising $N$ systematically 
yields more accurate spectra for both, Matsubara and Keldysh data, because in this case the requirements for the identity theorem of complex analysis which
guarantees uniqueness are approached systematically. However, the convergence
is always exponentially slow, as pointed out in the discussion of the dependence of the
singular-value distributions on $N$.

\subsubsection{Asymmetric triangle}
We next investigate 

\begin{equation}
A_\text{triag}(\omega) := \frac{1}{2}\cdot
\begin{cases}
\omega - 1, & \text{if } |\omega-2|<1, \\
0, & \text{else}
\end{cases}
\end{equation}
as an example for spectra which are not particle-hole symmetric and for
which the convergence of the inverse Fourier transform is more rapid than for
the rectangular case.
The variance is again set to $\epsilon^2 = 6\cdot 10^{-14}$.
\begin{figure}
\begin{center}
\includegraphics[width=\linewidth]{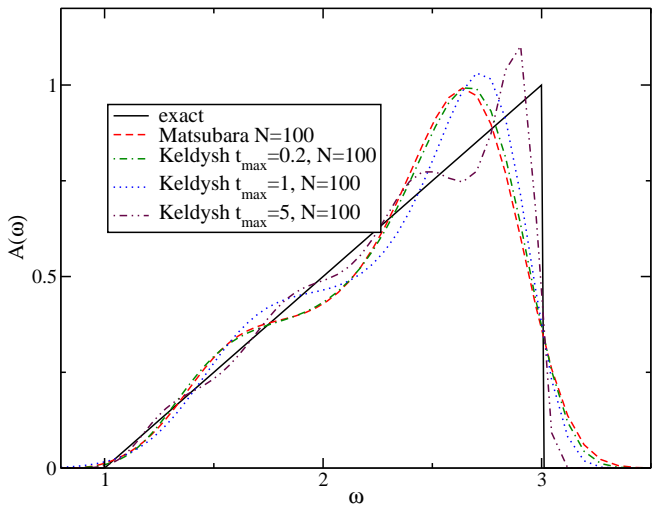}
\end{center}
\caption{(color online) Comparison for the triangular spectrum.}
\label{fig:dreieck}
\end{figure}
Results for this scenario are shown in Fig.~\ref{fig:dreieck}.
Judged by the peak position and the steepness at the discontinuity, even the
$t_\text{max}=0.2$ Keldysh spectrum is already slightly better than the Matsubara 
spectrum.
This indicates that real-time data are good for resolving {\it high-energy
features}, since the triangle is shifted to relatively high energies.
As in the case of the rectangular shape, the spectrum improves as $t_\text{max}$ is increased.

\subsubsection{Multiple peaks}

As a next step towards more realistic situations we turn to a spectrum with a
sharp resonance at $\omega=0$ and two side bands.
We model this  by superimposing Gaussians,
\begin{equation}
A_\text{peak}(\omega) :=
\sum_{\alpha=\pm 1}c_\text{sb} g_{\sigma_\text{sb},\alpha\Omega}(\omega)
+ c_\text{res} g_{\sigma_\text{res},0}(\omega),
\end{equation}
with $c_\text{res} = 0.1$, $c_\text{sb}=0.45$, $\Omega = 2.0$, 
$\sigma_\text{sb}=0.5$, $\sigma_\text{res}=0.05$;
$g_{\sigma,X}(x) :=
\frac{1}{\sqrt{2\pi\sigma^2}}\exp\left(-\frac{(x-X)^2}{2\sigma^2}\right)$.

\begin{figure}
\begin{center}
\includegraphics[width=\linewidth]{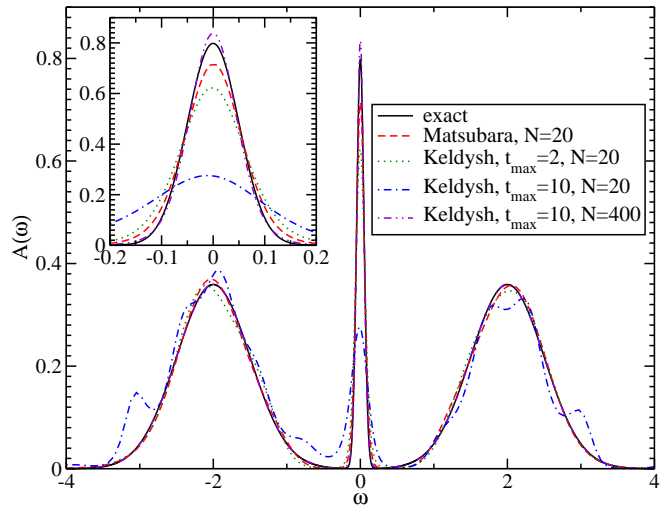}
\end{center}
\caption{(color online) Comparison for the three-peak spectrum.}
\label{fig:peakstest}
\end{figure}

As one can see in Fig.~\ref{fig:peakstest},  real-time data on relatively short contours 
($t_\text{max}=2$, $10$) are not particularly useful, 
when sharp \emph{low-frequency} features such as the 
given ``quasi-particle peak" should be extracted. 
The Matsubara data and the short-$t_\text{max}$ real-time data yield side bands of 
similar quality, while the imaginary-time data produce a comparable or even better reconstruction
of the sharp resonance.
Increasing $t_\text{max}$  mainly improves the resolution of smooth 
high-energy features ($t_\text{max}=10$, $N=400$).
We also observe that using $n=N/4=5$ equidistant
time steps on the $t_\text{max}=2$ real contour yields a better spectrum than a similar 
data set on the $t_\text{max}=10$ contour. This finding is 
compatible with the earlier suggestion that as a function 
of $N$, a full establishment of the singular value plateau is crucial. 
However, increasing $N$ at fixed $t_\text{max}$ involves only a polynomial increase of the computational 
effort, so we can always assume that $N$ is sufficiently large for the plateau to be fully developed.

Considering a broader selection of points from $G^\lceil(\tau,t)$ data, we did
not find an improvement of quality of the spectral function  compared to
pure Keldysh Green's functions.

\subsection{Comparison to direct inversion}
\label{inversion}

\begin{figure}
\includegraphics[width=\linewidth]{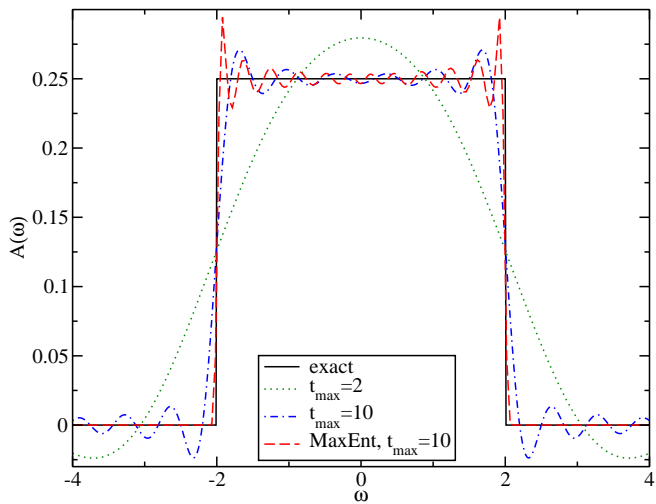}
\caption{(color online) Spectra obtained from retarded Green's function on the Keldysh contour
using a truncated Laplace transform ($t_\text{max}=2$ and $t_\text{max}=10$) and
comparison to the exact result. The $t_\text{max}=10$ Maximum Entropy result is
included for comparison. See Fig.~\ref{fig:rect} for a more detailed
comparison with Maximum Entropy data. }
\label{fig:directSharp}
\end{figure}

The maximum entropy deconvolution is superior to direct
inversion of the Fourier transform for finite Keldysh branches. To demonstrate this, 
we apply the usual rotation in Keldysh space and consider the retarded Green's 
function 
defined in Eq.~(\ref{retarded}).
In contrast to Eq.~\eqref{eq:kernelKeldysh},
no Fermi factor appears in the transform from the spectral function here,
\begin{equation}
\imag G^\text{ret}(t-t') = \int_{-\infty}^\infty\Dfrtl\omega
\euler{-\imag\omega(t-t')}A(\omega),
\quad t > t'.
\end{equation}
In the limit of a Keldysh contour of infinite length, a Laplace transform restores the
spectral function 
\begin{equation}
A(\omega) = -\Im \frac{1}{\pi} \int_0^\infty \Dfrtl t  \euler{+\imag\omega t} G^\text{ret}(t) .
\end{equation}
For our contour of finite lenght $t_\text{max}$ a straightforward approximation to the
deconvolution problem involves a truncation of the Laplace integral at time
$t_\text{max}$.
Figure \ref{fig:directSharp} shows spectra resulting from this method for
the rectangular test spectrum and contour lengths $t_\text{max}=2$ and $t_\text{max}=10$.

Comparison to the finite-$\epsilon$ maximum entropy results from Keldysh
propagators in
Fig.~\ref{fig:rect} shows that the sharp edge is not resolved as well. Furthermore,
unphysical regions of negative spectral weight appear in the case of the truncated
Laplace transform, whereas this is avoided by construction in the maximum entropy  
approach. 
The amplitude of the ringing oscillations seems to be almost the same as in the maximum entropy case, except near the band-edge, where
the truncation involves an arbitrary smoothening of sharp structures. 
Different cutoff procedures for the time integral would yield different smoothenings of the
spectrum. 

\section{Pad\'e approximation}
\label{Pade}

As a second alternative to the Maximum Entropy approach, we consider the Pad\'e method,
which in equilibrium situations is often superior to the Maximum Entropy method for the analytic continuation of
data without, or with very little stochastic noise.
 In particular for the analytical continuation of low-temperature Matsubara data, it is
known to be rather precise at low frequencies \cite{Otsuki2007}.

The Pad\'e approximant is constructed as follows: We assume that the values of the Green's function 
are known on a set of $N$ points $z_n$ in the complex frequency plane. Conventionally, the $z_n$ are
given by the Matsubara frequencies $\imag \omega_n$ ($\omega_n=(2n+1)\pi/\beta$). The Green function is interpolated with a 
rational function $C_N(z)$ in the form of a continued fraction, 
\begin{equation}
C_N(z) = \frac{a_1}{1+}\frac{a_2(z-z_1)}{1+}\cdots\frac{a_N(z-z_{N-1})}{1},
\end{equation}
where 
the coefficients $a_i$ of
the continued fraction 
are computed using a simple recursion formula \cite{Vidberg77}.
 
In the present paper we are in particular interested in the non-equilibrium situation,
i.e.~the determination of a time-dependent spectral function $A(t,\omega)$
from the real-time retarded Green's function on some finite time interval $[t,t+\Delta t]$.
For a set of points $z_n$ in the complex plane, we first compute an approximation $\tilde G(t, z_n)$ of the Fourier transformed $G^\text{ret}$,
\begin{equation}
\tilde G(t, z_n) := \int_t^{t+\Delta t} G^\text{ret}(t',t) \euler{\imag z_n (t'-t)}\Dfrtl{t'}.
\end{equation}
This approximation is good if the imaginary part of $z_n$ is sufficiently large 
($\text{Im}\, z_n \gg 1/\Delta t$).
One can then construct a
Pad\'e approximant to the function $\tilde G(t, z)$ which tends to suppress the
artificial oscillatory behaviour of spectral functions obtained from direct
Fourier transforms (corresponding to $z\to \omega +\imag 0^+$). 

\begin{figure}
\includegraphics[width=\linewidth]{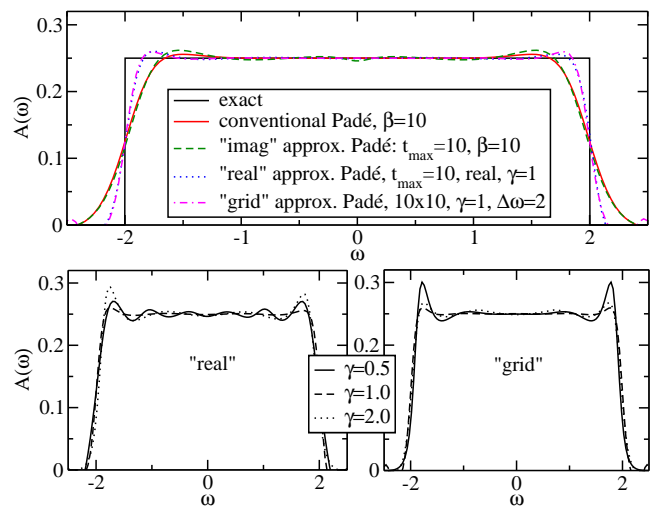}
\caption{(color online) Approximation of the rectangular spectrum with the Pad\'e approach, either exact or for 
$t_\text{max}=10$. See Figs.~\ref{fig:rect} and \ref{fig:directSharp} for comparison with Maximum Entropy and the truncated Laplace transform.
}
\label{fig:rectanglepade}
\end{figure}

For the approximate Pad\'e method we consider the following three variants: 
the points $z_n$ are
\begin{itemize}
\item
``imag": fermionic Matsubara frequencies
$z_n = \imag\omega_n$=$\imag (2n+1)\pi/\beta$, 
with some arbitrarily chosen $\beta$,
\item
``real": real frequencies with some imaginary offset $\imag \omegaii$, 
i.e.~$z_n = n\Delta\omega + \imag\omegaii$,
\item
``grid": arranged on a $N_\text{real}\times N_\text{imag}$ grid, where 
$z_{n_\text{real}n_\text{imag}} = \omega^{(r)}_{n_\text{real}} + \imag\omega^{(i)}_{n_\text{imag}}$. The lattice is chosen to be equidistant, with 
$\omega^{(r)}_{n_\text{real}}-\omega^{(r)}_{n_\text{real}-1} = \Delta\omega$ and $\omega^{(i)}_{n_\text{imag}} - \omega^{(i)}_{n_\text{imag}-1} =
\omega^{(i)}_1=:  \omegaii$.
\end{itemize}

Figure \ref{fig:rectanglepade} shows spectra obtained from these different
Pad\'e variants, when applied to the rectangular test spectrum.
For the approximate Pad\'e versions we always use the Keldysh contour length $t_\text{max}=10$, as well as 100 data points as input for any
of the Pad\'e procedures.
It is interesting to note that, as compared to the truncated Laplace transform and the Maximum Entropy approach, the amplitude of the oscillations is significantly smaller 
for the conventional Pad\'e. However, the resolution of the jump is clearly limited. As $\beta$ is decreased for the Matsubara frequencies, a slight
    increase of the slope can be observed.

Let us now discuss the approximate Pad\'e method for a time interval length $t_\text{max}=10$. By construction it is closely related to the truncated
Laplace transform, whose results are shown in Fig.~\ref{fig:directSharp}. Nevertheless, features of the original Pad\'e approach are also inherited, depending on how the  
$z_n$ are chosen. In fact, the ``imag" solution is much closer to the conventional Pad\'e solution than to the respective truncated Laplace transform. The undesirable
oscillations have been smoothened at the price of a reduced resolution at higher frequencies, while some of the oscillations from the truncated Laplace transform
survive. For $\omegaii=1$, the ``real" and ``grid" solutions are almost identical. These results are  
closer to the exact solution than the truncated Laplace transform, even near the jump,  and also exhibit smaller amplitude oscillations. Unfortunately, as shown in the lower panels, the rather unsystematic behaviour as a function of $\omegaii$ makes it difficult to determine the optimal value of $\omegaii$ a priori. We also note that in the ``real" variant, the truncated Laplace solution is recovered for $\omegaii=0.5$, since in this case all the $z_n$ are close to the real axis.

\section{Application to Dynamical Mean Field Results}
\label{application}

\subsection{Equilibrium spectra}

As a 
realistic application, we analyze data obtained for the Kondo lattice model (bandwidth $4$, $\beta=50$, $J=1.5$, $3.0$) within 
dynamical mean field theory in Ref.~\cite{Werner12kondo}, using the non-crossing approximation (NCA) as impurity solver. 
We do not want to discuss here the physics of the Kondo lattice
model, and to
what extent 
certain
high energy features in the spectral function may be an
artifact
of the
NCA. We merely use the results of Ref.~\cite{Werner12kondo}
as a nontrivial example involving low-energy quasi-particle peaks and high-energy
satellites. 

In order to apply the Maximum Entropy Method (MEM), we again assume a uniform error $\epsilon$ and
set offdiagonal entries of the covariance to zero. This seems justified as long
as the systematic numerical error of the simulation data is significantly smaller than $\epsilon$.
In practice, the value of $\epsilon$ can be rather easily determined by trial
and error: if $\epsilon$ is chosen too small, the MEM ceases to
converge, whereas if it is chosen too large, the data lack sufficient information.
In the following calculations, $\epsilon$ is chosen slightly larger than the value at which the MEM
breaks down. This moves the systematic numerical
errors just into the $\epsilon$-range of the Gaussian statistical errors
which are formally assumed by the Maximum Entropy procedure.
As input data we consider the Matsubara Green's functions $G^M(\tau)$ and the retarded
(equilibrium) Green's functions $G^\text{ret}(t-t')$, as well as data sets containing 
information from both Matsubara and retarded Green's functions.

\begin{figure}
\includegraphics[width=\linewidth]{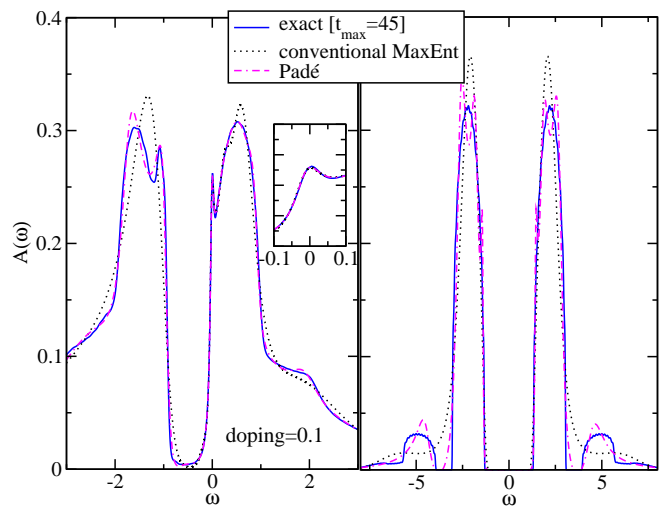}
\caption{(color online) Comparison of Pad\'e and conventional MEM spectra for a Kondo
lattice model (bandwidth $4$, inverse temperature $\beta=50$),  in the Fermi-liquid regime ($J=1.5$, left) and Kondo insulating regime  ($J=3.0$, right).
}
\label{fig:compaNCAPM}
\end{figure}

The NCA data considered here are noise-free low temperature data, and thus ideally suited for the Pad\'e method. In
Fig.~\ref{fig:compaNCAPM} we 
compare MEM, Pad\'e, and exact spectra
 for a half-filled
Kondo-insulator, 
and a doped heavy-Fermi liquid solution. The ``exact"
spectra have been obtained from a truncated Laplace transformation with large
$t_\text{max}\approx 45$. We see that both Pad\'e and Matsubara MEM
reproduce the low-energy features and the gap-edges very well, but the higher
energy structures cannot be accurately resolved. 

\begin{figure}
\includegraphics[width=\linewidth]{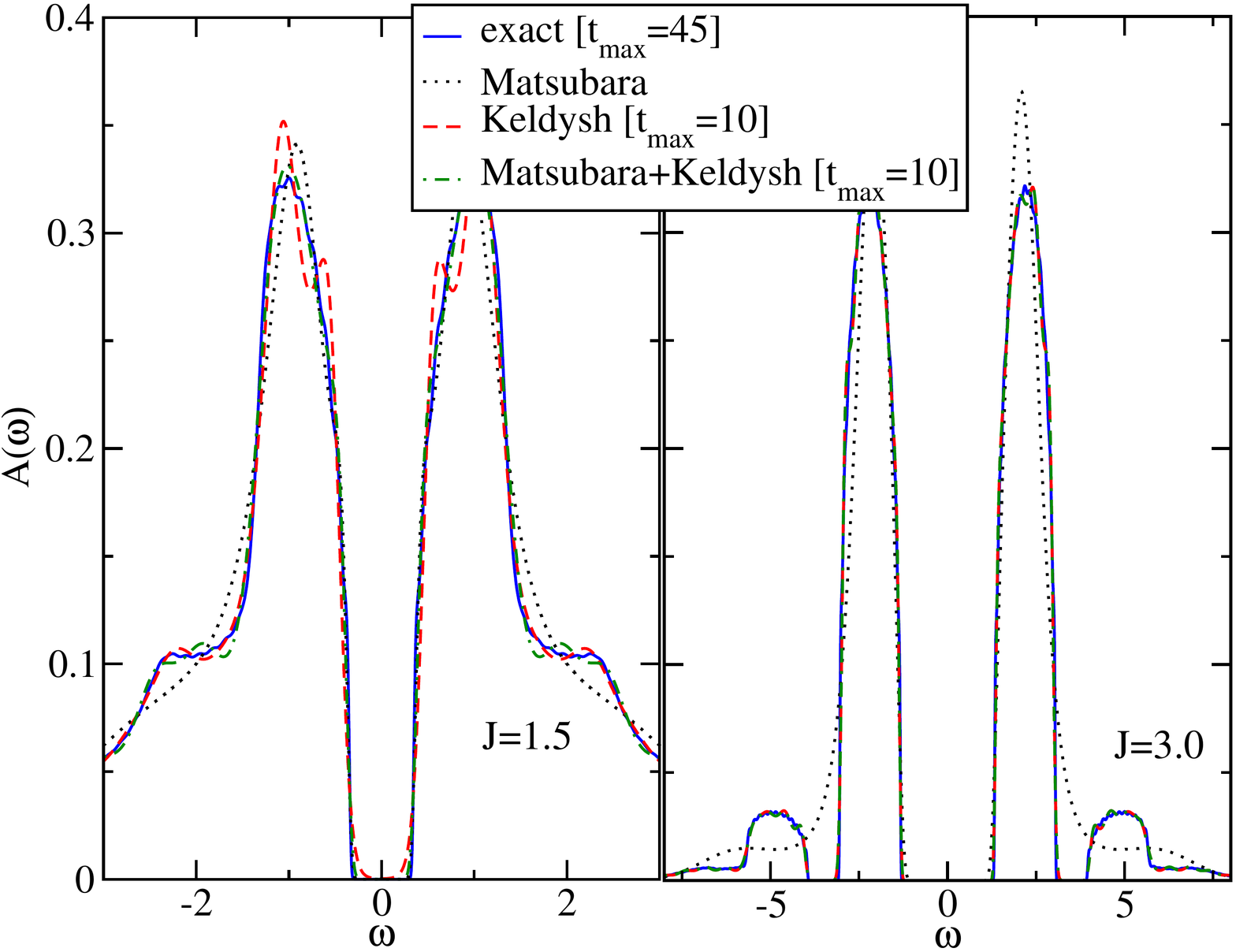}
\includegraphics[width=\linewidth]{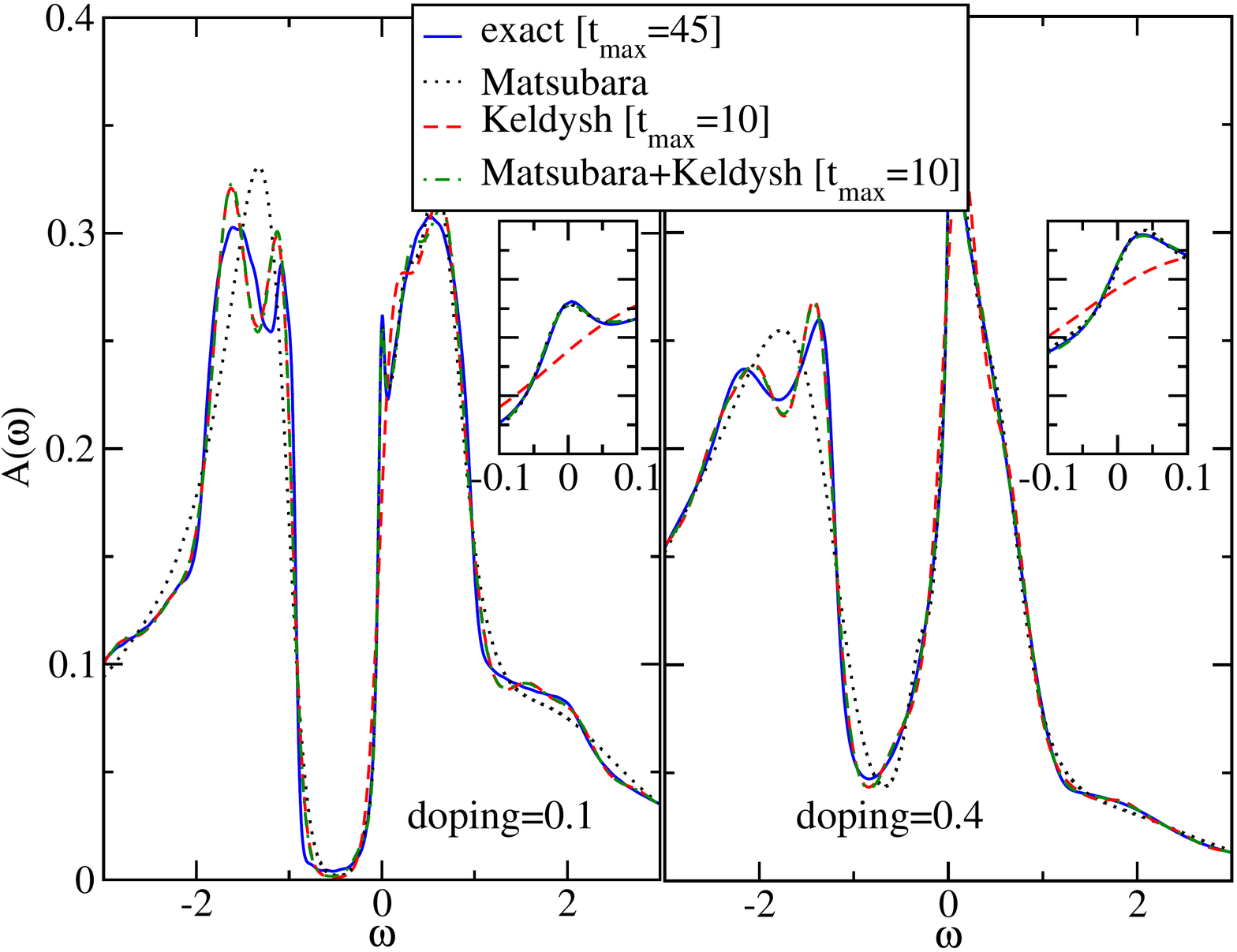}
\caption{
(color online) Comparison of MEM solutions for the insulating  Kondo lattice model (top, $J=1.5$ and $3.0$)
and the doped heavy Fermi liquid regime (bottom, $J=1.5$).  
A real-time contour length $t_\text{max}=10$ is sufficient to
extract the relevant features.
}
\label{fig:compaNCAinsu}
\end{figure}
In Fig.~\ref{fig:compaNCAinsu} 
we show in addition to the exact and Matsubara MEM spectra the results from a
MEM analytical continuation of real-time data with $t_\text{max}=10$ and of a
MEM continuation involving both the real-time and the imaginary-time data.
The real-time MEM spectra produce the correct high-frequency behavior, and at
least qualitatively correct structures in the lower band of the doped system,
but they fail to reproduce the quasi-particle peaks. By adding also the imaginary-time data, we can accurately resolve the low-energy behavior (see insets), in addition to the high-energy structures. The main challenge in the equilibrium case thus remains the calculation of spectral features in the intermediate-energy range, where the convergence to the exact result with increasing $t_\text{max}$ is relatively slow.

\begin{figure}
\includegraphics[width=\linewidth]{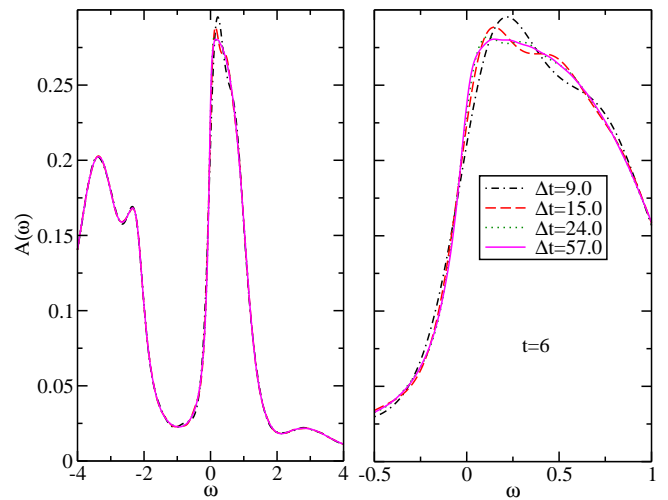}
\caption{(color online)  
Convergence of $A(\omega, t=6)$ for different time-intervals $[t,t+\Delta t]$. 
}
\label{fig:a_t_maxent}
\end{figure}

\subsection{Time-dependent spectra}

To test the ability of the MEM approach to extract time-dependent spectra, we
again consider results from Ref.~\onlinecite{Werner12kondo}. These spectra
correspond to a perturbed, doped Kondo-lattice model which dissipates energy
to a heat-bath, and thereby evolves from the high-temperature local moment regime into the
low-temperature heavy-Fermi liquid regime. Again, we are not interested in the physics here,
which has been discussed in Ref.~\cite{Werner12kondo}, but 
just in the 
quality of the spectral functions 
that can be 
extracted from the retarded Green functions.
In particular, we want to investigate the constraints on  $t_\text{max}$ for the extraction of a time-dependent
spectral function at time $t$ ($t <t_\text{max}$), since most computational
techniques are limited to short time contours. 

The upper panels of Fig.~\ref{fig:a_t_maxent} show the spectral function
$A(\omega,t)$ at a relatively short time $t=6$ after the perturbation. As can
be seen, $A(\omega,t)$ is already positive over the whole relevant frequency
range. The different curves in the figure correspond to different
time windows $[t,t+\Delta t]$ used in the MEM analytical continuation. The
result for the largest $\Delta t$ (dash-dotted curve) can be considered the exact
result. We learn from these results that the high-frequency part converges
very quickly with $\Delta t$, while we need at least an interval of length $\Delta t=24$ to get a reasonably accurate result at
low energies. 
(Since this calculation is a real nonequilibrium application, we cannot resort to Matsubara data to fix the low-energy part of the spectral function.)

A relevant question is how the MEM approach performs compared to direct
Fourier transformation on an identical time interval. 
While both
the MEM method and the direct Fourier transformation have difficulties
reproducing the correct low-energy peak, the MEM spectra are slightly closer to the
correct result. 

\begin{figure*}
{ \includegraphics[width=0.48\linewidth]{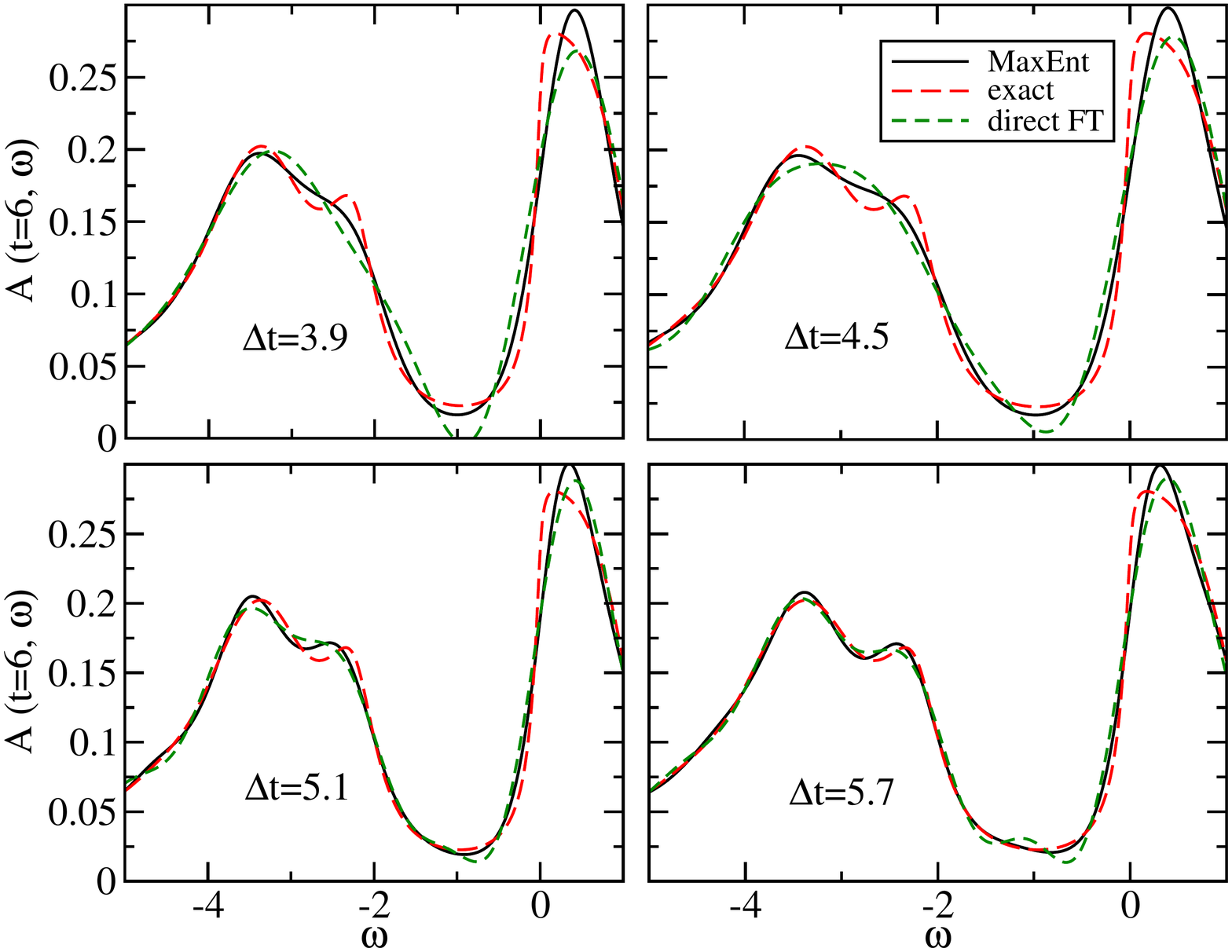}
}
{ \includegraphics[width=0.48\linewidth]{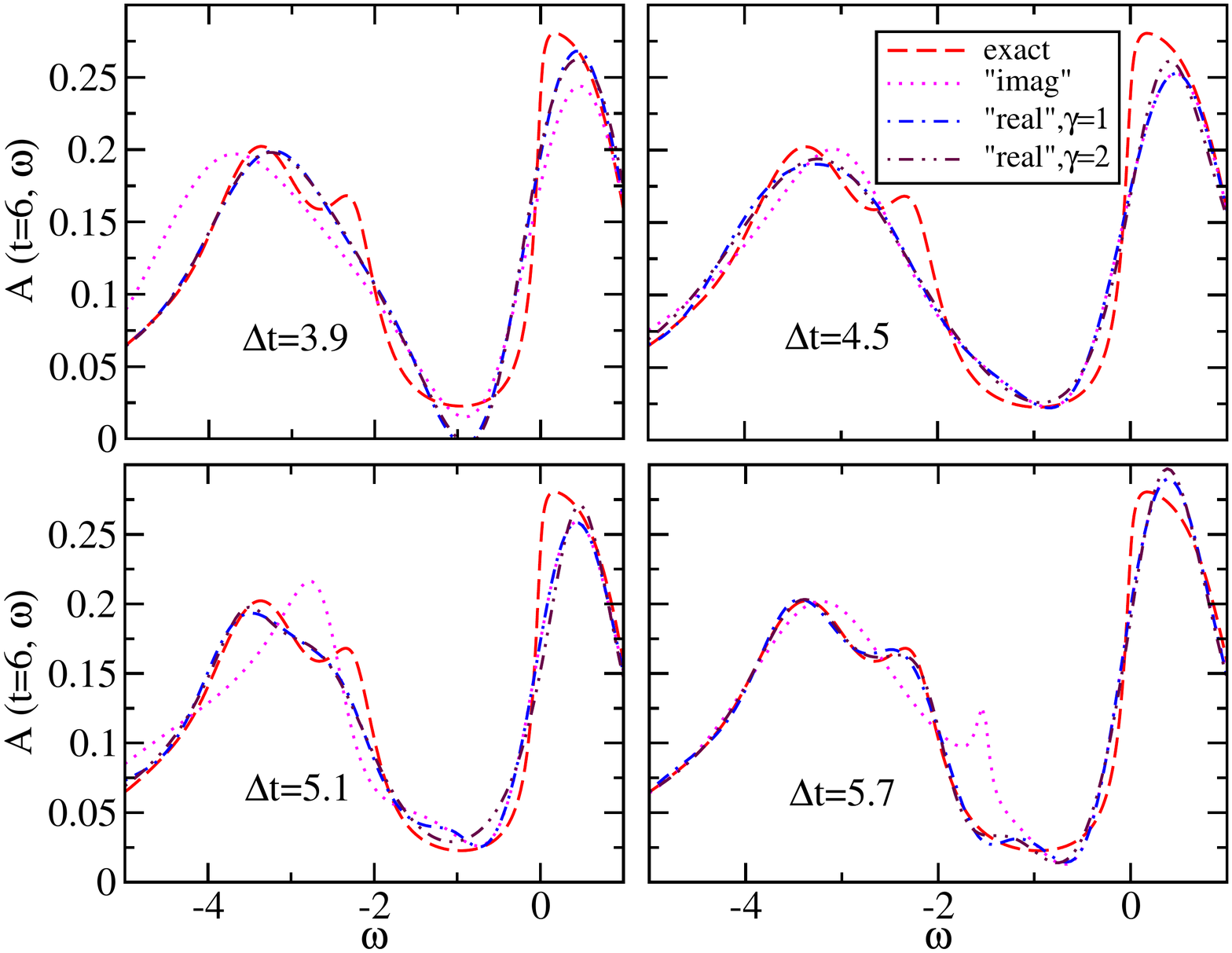}
}
\caption{Convergence of non-equilibrium $A(\omega,t=6)$ estimates for different short
real-time intervals. In the case of the ``imag" modified Pad\'e, we set $\beta=20$ and used 100 data points. For the ``real" modified Pad\'e results, we
used 100 equidistant data points for $\Re z_n$ on the interval $[-10,10]$.
}
\label{fig:shortintvsnoneq}
\end{figure*}

\subsection{Convergence of high-energy features} 

While the convergence of the MEM result for pure real-time Green's function data as a function
of $\Delta t$ is slow for the quasi-particle 
peak, it is interesting to see how the higher-energy features of the
non-equilibrium spectral function converge to the exact result. 

Figure \ref{fig:shortintvsnoneq} shows a comparison of the MEM result to the direct Fourier transform and the modified Pad\'e method (discussed in Sec.~\ref{Pade}) for short time intervals with lengths $\Delta t \in [3.9, 5.7]$.
For $\Delta t=5.7$, MEM and direct Fourier transformation produce the correct solution, except in the low-energy region $-0.5\lesssim \omega \lesssim 0.5$, the modified Pad\'e method for the imaginary time interval yields spurious results. 
For shorter interval lengths, the MEM appears to be slightly superior to both, direct Fourier transform and modified Pad\'e.
All qualitative features of the spectrum, including the double-peak structure in the lower band start to become visible in the MEM solution at 
$\Delta t=4.5$, whereas the emergence of similar structures in the direct FT spectrum at $\Delta t=5.1$ may still be interpreted as part of the overall oscillatory behaviour
of that solution. One may thus argue that the MEM reduces the necessary interval length from $\Delta t = 5.7$ to $\Delta t=4.5$ and thus allows to compute time-dependent spectra up so slightly longer times.  
Depending on the application, this could safe considerable effort, since numerical algorithms typically scale badly (at least as the third power) with the length of the contour. 
The effect of the modified Pad\'e for $z_n$ on parallel lines to the real axis can to a large extent be interpreted as a smoothening procedure for the direct FT. 
In particular, the double-peak structure in the direct FT solution at $\Delta t=5.1$ is strongly suppressed, but still slightly visible. However, the peaks are less 
pronounced than those of the MEM solution at $\Delta t=4.5$.
The behaviour of the ``grid" variant is very similar to the ``real" variant, except that the ``grid" choice cannot resolve the double-peak structure 
of the lower band (not shown). When the number of real frequency points is increased, the ``grid" result converges to the ``real" result. Due to numerical instabilities of the
Pad\'e procedure, one is however typically limited to less than 400 grid points.

\section{Conclusion}
\label{conclusion}
We analyzed the usefulness of Keldysh real-time Green's
function data for the computation of equilibrium spectral functions of interacting
quantum many-body systems.
The ill-posed nature of the inversion corresponding to the
determination of the spectral function is reflected in the distribution of
singular values of the respective continuation kernel. The
conventional Wick rotation of Matsubara data yields an exponentially
decaying singular value distribution. We found that including a finite real-time
branch to the imaginary-time contour adds a plateau to this distribution, which broadens as the length of 
the real-time contour is increased. This plateau significantly alleviates the inversion even for small
contour lengths. A further analysis showed that the main gain is 
more precise high-frequency information of the spectral function. For short
contours, the real-time data however provide rather crude results at low
frequencies. In order to obtain more accurate estimates of low-energy features such as quasi-particle peaks,
it is therefore necessary to include data from the Matsubara branch.
With the small frequencies well covered by the Matsubara branch and the high
frequencies well covered by a short Keldysh branch, the biggest challenge remains the resolution at intermediate energies.

Our analysis of NCA spectra for the Kondo-lattice model 
suggests that in order to resolve intermediate energies reasonably well, 
the length of the real-time contour has to be at least $t_\text{max}=10$ (in units where the bandwidth is $4$). In real-time Monte Carlo simulations, it is difficult to reach these times, at least in parameter regimes where low-order perturbation theory is not reliable. It thus appears that the exponential increase in the computational cost with increasing $t_\text{max}$ in real-time Monte Carlo schemes outweighs the benefits from an alleviated MEM analytical continuation. 
Nevertheless, the MEM analytical continuation of Keldysh Green functions can be a useful and superior alternative to
Pad\'e approximants for semi-analytic methods such as higher order strong-coupling perturbation theory. 
These methods are very useful in certain parameter 
regimes, such as elevated temperature, or intermediate to strong interactions. While
their implementation on the real-frequency axis is challenging, calculations on the Keldysh contour scale polynomially with $t_\text{max}$ and the contour-lengths needed for reliable MEM analytical continuation can be reached.  

In the calculation of non-equilibrium spectra, it is no longer possible to include the Matsubara
branch into the Maximum Entropy procedure. 
As a consequence, the resolution of low-energy features of the spectrum 
worsens dramatically if only short-time data are available.
Nevertheless, a comparison to alternative, more
straightforward techniques, i.e.~direct Laplace transform and the modified
Pad\'e approach, showed that the MEM yields slightly more reliable solutions for these
spectra. The relevant structures could be established for somewhat shorter time intervals than
for the extended Pad\'e approaches.

On a conceptual level, the MEM is superior to the direct Laplace transform and the generalized Pad\'e approach. 
In practical applications, however, because 
the advantage of a MEM continuation appears to be rather subtle, it will 
very much depend on the details of the utilized many-body approach whether an
application of MEM is worth its effort. In any case, the comparison of 
different approaches can help in deciding which features 
of a non-equilibrium spectrum are trustworthy.

\acknowledgements
We acknowledge funding from the Deutsche
Forschungsgemeinschaft (DFG)
via SFB 602, SNF grant PP0022-118866 and FP7/ERC starting
grant No. 278023. This work also profited from the GoeGrid initiative and the Gesellschaft f\"ur
Wissenschaftliche Datenverarbeitung G\"ottingen (GWDG).


\begin{thebibliography}{99}

\bibitem{Eckstein09} M. Eckstein, M. Kollar and P. Werner, Phys. Rev. Lett. {\bf 103}, 056403 (2009).
\bibitem{Eckstein10} M. Eckstein, M. Kollar and P. Werner, Phys. Rev. B {\bf 81}, 115131 (2010).
\bibitem{Werner10} P. Werner, T. Oka, M. Eckstein, and A. J. Millis, Phys. Rev. B {\bf 81}, 035108 (2010).
\bibitem{Rubtsov05} A. N. Rubtsov, V. V. Savkin, and A. I. Lichtenstein, Phys. Rev. B {\bf 72}, 035122 (2005).
\bibitem{Werner06} P. Werner, A. Comanac, L. de' Medici, M. Troyer, and A. J. Millis, Phys. Rev. Lett. {\bf 97}, 076405 (2006). 
\bibitem{Muehlbacher08} L. M\"uhlbacher and E. Rabani, Phys. Rev. Lett. {\bf 100}, 176403 (2008).
\bibitem{Werner09} P. Werner, T. Oka, and A. J. Millis, Phys. Rev. B {\bf 79}, 035320 (2009).
\bibitem{Tsuji2012afm} N. Tsuji, M. Eckstein and P. Werner,  arXiv:1210.0133.
\bibitem{Eckstein10nca} M.~Eckstein and P.~Werner, Phys. Rev. B {\bf 82}, 115115 (2010).
\bibitem{Gull2010} E. Gull, D. Reichmann and A. J. Millis, Phys. Rev. B {\bf 84}, 085134 (2011).
\bibitem{Eckstein2009} M. Eckstein, M. Kollar and P. Werner, Phys. Rev. B {\bf 81}, 115131 (2010).
\bibitem{Eckstein2011pump} M. Eckstein and P. Werner, Phys. Rev. B {\bf 84}, 035122.
\bibitem{Tsuji2012} N. Tsuji, T. Oka, H. Aoki and P. Werner, Phys. Rev. B {\bf 85}, 155124 (2012).
\bibitem{Eckstein2008} M. Eckstein and M. Kollar, Phys. Rev. B {\bf 78}, 245113 (2008).
\bibitem{Freericks2009} J. K. Freericks, H. R. Krishnamurthy, and Th. Pruschke, Phys. Rev. Lett. {\bf 102}, 136401 (2009). 

\bibitem{Schwinger61}J.~Schwinger, J.~Math.~Phys.~{\bf 2}, 407 (1961)
  \bibitem{keldyshintro}
    For an introduction into the Keldysh formalism, see, e.g., 
H.~Haug and A.-P. Jauho,
\newblock {\em Quantum Kinetics in Transport and Optics of Semiconductors},
\newblock Springer, Berlin, 2nd edition, 2008.
\bibitem{jaynes57}E. T. Jaynes, Phys. Rev. {\bf 106}, 620 (1957)
\bibitem{jarrell_review}M.~Jarrell, J.~E.~Gubernatis, Physics Reports {\bf
269}, 133 (1996).
\bibitem{maxent_buch}N. Wu, ``The Maximum Entropy Method", 162 (1997).
\bibitem{bryan}R.~K.~Bryan, Eur.~Biophys.~J.~{\bf 18}, 165 (1990).
\bibitem{Jeffreys46}H.~Jeffreys,
Proceedings of the Royal Society of London, Series A {\bf 186}, No.~1007, 453
(1946)
\bibitem{jarrell} M.~Jarrell, A.~Macridin, K.~Mikelsons, and 
D.G.S.P.~Doluweera, in \emph{Lectures on the Physics of Strongly Correlated
Systems XII}, AIP Conference Proc. {\bf 1014}, A. Avella and F. Mancini
(Eds.), 34 (2008)
\bibitem{Otsuki2007} J. Otsuki, H. Kusunose, P. Werner, and Y. Kuramoto, J. Phys. Soc. Jpn. {\bf 76}, 114707 (2007).
\bibitem{Vidberg77} H. J. Vidberg and J. W. Serene, J. Low Temp. Phys. {\bf 29} (1977) 179.
\bibitem{Werner12kondo} P. Werner and M. Eckstein, Phys. Rev. B {\bf 86}, 045119 (2012). 
\end{thebibliography}
\end{document}